\documentclass[lettersize, journal]{IEEEtran}  

\usepackage{graphics} 
\usepackage{epsfig} 
\usepackage{mathptmx} 
\usepackage{times} 
\usepackage{amsmath} 
\usepackage{amsthm}
\usepackage{amssymb}  
\usepackage{subcaption}
\usepackage{graphicx}
\usepackage{algorithm}
\usepackage[colorlinks,allcolors=blue]{hyperref}
\usepackage{cite}
\usepackage{lipsum}
\usepackage{cuted}
\usepackage{indentfirst}
\usepackage{multirow}
\title{\LARGE \bf
Event-Based Adaptive Koopman Framework for Optic Flow-Guided Landing on Moving Platforms}
\author{Bazeela Banday, Chandan Kumar Sah and Jishnu Keshavan%
\thanks{ The authors are with the Department of Mechanical Engineering, Indian Institute of Science. \tt\small \{bazeelab, chandanks, kjishnu\}@iisc.ac.in}
}

\begin{document}

\maketitle
\thispagestyle{empty}
\pagestyle{empty}

\begin{abstract}
This paper presents an optic flow-guided approach for achieving soft landings by resource-constrained unmanned aerial vehicles (UAVs) on dynamic platforms. An offline data-driven linear model based on Koopman operator theory is developed to describe the underlying (nonlinear) dynamics of optic flow output obtained from a single monocular camera that maps to vehicle acceleration as the control input. Moreover, a novel adaptation scheme within the Koopman framework is introduced online to handle uncertainties such as unknown platform motion and ground effect, which exert a significant influence during the terminal stage of the descent process. Further, to minimize computational overhead, an event-based adaptation trigger is incorporated into an event-driven Model Predictive Control (MPC) strategy to regulate optic flow and track a desired reference. A detailed convergence analysis ensures global convergence of the tracking error to a uniform ultimate bound while ensuring Zeno-free behavior. Simulation results demonstrate the algorithm’s robustness and effectiveness in landing on dynamic platforms under ground effect and sensor noise, which compares favorably to non-adaptive event-triggered and time-triggered adaptive schemes.


\end{abstract}

\section{INTRODUCTION}\label{sec1}

Achieving smooth, autonomous landings on a moving platform with minimal sensor reliance is a significant technical challenge for resource-constrained Unmanned Air Vehicles (UAVs). While technologies like lidar~\cite{dougherty2014laser} and stereo cameras offer high accuracy, their use in UAVs is limited by weight, computational demands, and power consumption~\cite{badrloo2022image}. Drawing inspiration from honeybees' ability to land without altitude or velocity data, recent research has focused on optic flow-based strategies for autonomous UAV landing on stationary platforms~\cite{srinivasan2000honeybees}. For instance, the study \cite{nabavi2022automatic} developed a monocular camera-based control strategy to estimate vertical distance for achieving smooth landing. However, this method's reliance on state estimation increases computational complexity and is prone to scaling issues due to noisy data~\cite{serra2016landing}. To overcome these limitations,~\cite{dupeyroux2021neuromorphic} propose a learning-based approach using a Spiking Neural Network (SNN) that directly maps optic flow to thrust, bypassing the design and implementation of an explicit controller. Yet, like other learning-based methods, this approach lacks robustness and struggles to generalize outside the training data, making it unreliable in the presence of uncertainties like the ground effect.

The problem of optic flow-guided autonomous landing on moving platforms is tackled in~\cite{xuan2020autonomous} using an infrared (IR) camera and a beacon to estimate the target's state. However, this method is restricted to platforms equipped with IR markers. The study in ~\cite{arif2023finite} enhances this strategy by fusing a monocular camera with an Inertial Measurement Unit (IMU) to estimate the platform's sinusoidal motion for landing. This approach, however, is limited by its reliance on the assumption of sinusoidal motion, making it less adaptable to other motion patterns. 

The dynamics of UAVs, with thrust/acceleration as input and optic flow as output, become increasingly complex near the ground or confined spaces due to nonlinear aerodynamic effects. These unmodeled dynamics complicate stability, particularly during soft landings on moving platforms. To tackle this, the current study relies on data-driven learning to model the relationship between optic flow and thrust in vertical motion. Using a Koopman operator-based framework, a global linear model is synthesized from experimental data, which simplifies control design and enables cost-effective optimal control. To this end, the Extended Dynamic Mode Decomposition (EDMD)~\cite{proctor2016dynamic} is invoked to generate an offline Koopman model using polynomial basis functions. However, EDMD struggles to generalize beyond the training data, and unmodeled dynamics from ground effects and platform oscillations lead to errors in the nominal Koopman model. To address this drawback, the study in~\cite{singh2024adaptive} proposes an online adaptation module using a neural network, however, its high computational cost is a significant concern. Instead, this study proposes developing an online adaptation law that updates the linear Koopman model parameters by minimizing the error between predicted and observed lifted states over an adaptation window.

While this proposed adaptation law is computationally efficient, using a monocular camera for optic flow computation can be resource-intensive for UAVs with limited onboard resources~\cite{7743546}. Event-triggered control (ETC), which updates only when necessary, offers a more efficient alternative to time-triggered control systems, reducing unnecessary updates and sensor measurements~\cite{6310015}. Past studies have demonstrated ETC's effectiveness in minimizing update frequency while maintaining performance~\cite{zhou2023event}. To address the limitations of existing approaches, an Event-Triggered Adaptation and Control (ETAC) scheme is proposed in this study that independently triggers adaptation and control updates based on their respective triggering mechanisms.

Thus, this work introduces a novel event-triggered Koopman-based data-driven approach for guaranteed soft vertical landing on an oscillating platform using a monocular camera's visual cues. The key contribution lies in synthesizing an event-triggered linear model incorporating an online adaptation algorithm to refine the nominal Koopman model in real-time. In particular, this study presents the first integration of a novel event-triggered adaptation and control policy within a linear Koopman framework, which ensures model and control updates occur only when necessary, thus reducing the computational cost during practical implementation. Detailed convergence analysis is used to establish global convergence of the optic flow error to a uniform ultimate bound while ensuring Zeno-free behavior. Simulation results are used to demonstrate the efficacy of the proposed scheme, including a favorable comparison with alternative event-triggered designs.

\section {Preliminaries and Problem Statement}\label{sec2}
This section formally defines the problem of soft vertical landing on a moving platform and the associated optic flow dynamics. A concise background of the Koopman operator theory is also provided, which is used to formulate the adaptation law for online model correction.

\subsection{Notation}
The sets \( \mathbb{R} \), \( \mathbb{R}^n \), and \( \mathbb{R}^{n \times m} \) represent real numbers, vectors of length $n$, and matrices of size $n \times m$. Non-negative real numbers are denoted by \( \mathbb{R}_{\geq 0} \). A positive definite or semi-definite matrix \( \boldsymbol{P} \) is denoted by \( \boldsymbol{P} > 0 \) or \( \boldsymbol{P} \geq 0 \), and the 2-norm of a vector \( x \in \mathbb{R}^n \) is \( \|x\| \). $\lambda_{min}\{Q\}$ and $\lambda_{max}\{Q\}$ are the minimum and maximum eigenvalues of $Q$ respectively. A function \( \eta : \mathbb{R}_{\geq 0} \to \mathbb{R}_{\geq 0} \) belongs to class \( \mathcal{K} \) if it is continuous, strictly increasing, and \( \eta(0) = 0 \). A function \( \kappa : \mathbb{R}_{\geq 0} \to \mathbb{R}_{\geq 0} \) is in class \( \mathcal{K}_{\infty} \) if it is in \( \mathcal{K} \) and \( \kappa(r) \to \infty \) as \( r \to \infty \).

\subsection{Optic flow dynamics of a Quadrotor}\label{sec2.1}
This work adopts the optic flow formulation from ~\cite{singhal2023constant}, which assumes that the center of the sensor frame coincides with the center of the body frame of the vehicle. Optic flow is induced by the relative motion of the UAV in the vertical direction, with output defined as $x(t) = {v (t)}/{h(t)}$
where $v(t)$ is the downward velocity of the UAV, $h(t)$ is the height above the landing surface, and $x(t)$ is the measured optic flow output. The complexities of autonomous UAV landing on a moving surface, particularly when the platform's motion is unknown, become significantly more challenging due to the vertical oscillations of the landing surface. These challenges are further amplified by environmental disturbances and variations in the platform's motion. 
The optic flow dynamics of a UAV for vertical landing on a moving platform in the presence of ground effect can be obtained from \cite{singhal2023constant} as:
\begin{eqnarray}
        \dot{x}(t) = - x^{2} (t) + \frac{\zeta(t) a}{h(t)} + \Delta_1(t) + \Delta_2(t),
\label{eq:GE MP}
\end{eqnarray}
where $\Delta_2$ is the additional uncertainty induced by the moving platform and given by ${-a_p}/{h(t)}$ and $a_p$ being the acceleration of the moving platform, $\zeta(t) \ge 1$ is the factor by which thrust changes in the presence of a ground effect, $\Delta_1(t) = (\zeta-1)g/h(t)$ is the perturbation term due to ground effect, $g$ is the acceleration due to gravity. Since $h(t) = h(0)e^{x(t)t}$, 
a smooth landing can be achieved by regulating optic flow output to a negative constant reference value $x_{ref}$.  

While (\ref{eq:GE MP}) describes the optic flow dynamics for landing on a vertically oscillating platform, it involves several uncertain parameters, including those dependent on vertical height and platform acceleration. To account for these uncertainties, we synthesize a data-driven linear Koopman model of observables that are polynomial functions of optic flow $x(t)$. Additionally, an online adaptation framework is designed to continuously refine the offline model to account for uncertainty arising from ground effects and platform motion.


\subsection{Preliminaries on Koopman Operator Theory}
The mathematical formulation of the Koopman operator theory and its applications to autonomous and nonautonomous systems has been extensively studied in the literature ~\cite{goswami2017global, sah2024real}. Hence, we only highlight the key concept of the Koopman operator theory pertinent to our application. Consider a discrete-time control-affine system of form
\begin{eqnarray}
    \label{eq:controlled_sytem}
    \boldsymbol{{x}}_{k+1} = \boldsymbol{f_0}(\boldsymbol{x}_k) + \sum_{i=1}^m \boldsymbol{f}_i(\boldsymbol{x}_k) \boldsymbol{u}_k,\; \boldsymbol{y}_k = \boldsymbol{h}(\boldsymbol{x}_k),
\end{eqnarray}
where $\boldsymbol{x}_k \in \mathbb{X} \subseteq \mathbb{R}^n $ is the state vector and $\boldsymbol{u} \in \mathbb{U} \subseteq \mathbb{R}^m$ is the control vector at $k^{th}$ timestep, $\boldsymbol{x}_{k+1}$ is the state vector at $(k+1)^{th}$ timestep and $\boldsymbol{f}_0$ is the drift vector field and $\boldsymbol{f}_i \forall i=1,..,m$ are the control vector fields and $\boldsymbol{h}$ is the output function. With a suitable choice of the observable functions $\boldsymbol{z} \in \mathbb{R}^q$, the control-affine system (\ref{eq:controlled_sytem}) attains a linear Koopman representation of the form~\cite{korda2018linear},
\begin{eqnarray}
    \label{eq:Koopman_lin}
    \boldsymbol{z}_{k+1} = \boldsymbol{A} \boldsymbol{z}_k + \boldsymbol{B} \boldsymbol{u}_k,\; \boldsymbol{x}_k = \boldsymbol{C} \boldsymbol{z}_k,
\end{eqnarray}
where $\boldsymbol{z}_k = \varphi(\boldsymbol{x}_k)$ is a nonlinear mapping from the finite-dimensional state space $\mathbb{X}$ to the finite-dimensional Hilbert space, $\boldsymbol{A} \in \mathbb{R}^{q \times q}$, $\boldsymbol{B} \in \mathbb{R}^{q \times m}$, $\boldsymbol{C} \in \mathbb{R}^{n \times q}$ is the projection matrix from lifted state to state-space. Following standard practice, the first $n$ elements of $z_k$ contain $x_k$, so that $C$ can be appropriately partitioned as $C=[I,0]$. 
 
\subsubsection{Extended Dynamic Mode Decomposition (EDMD)}
 Extended Dynamic Mode Decomposition (EDMD) provides a data-driven algorithm to approximate a finite-dimensional Koopman operator from input-output data. It uses a dictionary of nonlinear functions of outputs $\varphi( \boldsymbol{x})$ that is assumed to span the Koopman invariant subspace in which the dynamics of the system evolves. The EDMD algorithm computes the finite-dimensional Koopman approximation from dataset $\mathcal{D} = \{\boldsymbol{{x}}_j, \boldsymbol{u}_j,\boldsymbol{{x}}_{j+1}\}, j = 1,2,...,N-1$ by solving the minimization problem, $\sum_{j=1}^N \| \boldsymbol{\varphi}(\boldsymbol{{x}})_{j+1} - \boldsymbol{A} \boldsymbol{\phi}(\boldsymbol{{x}}_j) - \boldsymbol{B} \boldsymbol{u}_j \|_2^2$. The linear Koopman model obtained after the optimization process is given by (\ref{eq:Koopman_lin}).

\subsubsection{Adaptive EDMD Formulation}
We now formulate the linear adaptive EDMD framework for an uncertain control-affine system dynamical system described by,
\begin{eqnarray}
\label{eq:uncertain}
    \boldsymbol{x}_{k+1} = \boldsymbol{f_0}(\boldsymbol{x}_k) + \boldsymbol{\tilde{f}_0}(\boldsymbol{x}_k)+ \sum_{i=1}^{m} (\boldsymbol{f}_i(\boldsymbol{x}_k) + \boldsymbol{\tilde{f}}_i(\boldsymbol{x}))u_{i,k},
\end{eqnarray}
where $\boldsymbol{\tilde{f}_0}$ and  $\boldsymbol{\tilde{f}_i}$ represent the perturbations to the drift and actuation terms, respectively. It has been shown in \cite{singh2024adaptive} that the uncertain system (\ref{eq:uncertain}) has the following Koopman representation provided that the unmodelled/uncertain dynamics also lies in Koopman invariant subspace spanned by the basis functions used for offline model learning. Thus, we have,
\begin{eqnarray}
\label{eq:uncertain_koop}
\boldsymbol{z}_{k{+}1} {=} (\boldsymbol{A} + \boldsymbol{\Delta A}_k)\boldsymbol{z}_{k} {+} (\boldsymbol{B} + \boldsymbol{\Delta B}_k)\boldsymbol{u}_k, \, \,\boldsymbol{x}_k {=} \boldsymbol{C}\boldsymbol{z}_k,
\end{eqnarray}
where $\boldsymbol{\Delta A}_k$ and $\boldsymbol{\Delta B}_k$ are the updates made to account for uncertainties in the system (\ref{eq:uncertain}). Here, $\boldsymbol{z}_{k+1}$ is the lifted state obtained after lifting the observed state $\boldsymbol{x}_{k+1}$ at $(k+1)^{th}$ time-step i.e, $\boldsymbol{z}_{k+1} = \varphi (\boldsymbol{x}_{k+1})$. Hence, the objective is to compute the updates $\boldsymbol{\Delta A}$ and $\boldsymbol{\Delta B}$ that minimize the difference between the observed and predicted lifted state. From (\ref{eq:Koopman_lin}) and (\ref{eq:uncertain_koop}), the objective function to be minimized is defined as,
\begin{eqnarray}
\label{eq:obj}
J_k {=} \min_{\boldsymbol{\Delta A_k, \Delta B_k}} \frac{1}{2}\boldsymbol{y}_k^{\top}\boldsymbol{y}_k,\, \boldsymbol{y}_k{=}\boldsymbol{\Delta z}_k {-} \boldsymbol{\Delta A}_k \boldsymbol{z}_{k-1} {-} \boldsymbol{\Delta B}_k \boldsymbol{u}_{k-1},
\end{eqnarray}
where $\boldsymbol{\Delta z}_k = \boldsymbol{z}_{k} - \boldsymbol{\hat{z}}_{k}$ represents the prediction error for the linear Koopman representation. We define the update law as,
\begin{eqnarray}
\label{eq:pseudo}
    \begin{bmatrix}
        \boldsymbol{\Delta A}_k & \boldsymbol{\Delta B}_k
    \end{bmatrix}
    = \boldsymbol{\Delta z}_k
    \left(
    \begin{bmatrix}
    \boldsymbol{z}_{k-1}^{\top} & \boldsymbol{u}_{k-1}^{\top}
    \end{bmatrix}^{\top}
    \right)
    ^{\dagger}.
\end{eqnarray}
where $(^{\dagger})$ is the right pseudo-inverse. We now have the following lemma.

\emph{Lemma 1: The time-triggered update law (\ref{eq:pseudo}) minimizes the objective function (\ref{eq:obj}}) with the minimum obtained as $J_{k,min} = \boldsymbol{e}_{k,min}^{\top} \boldsymbol{e}_{k,min}$,
where $\boldsymbol{e}_{k,min} {=} 
\boldsymbol{\Delta z}_k [\boldsymbol{I} -
\{
\begin{bmatrix}
\boldsymbol{z}_{k-1}^{\top} & \boldsymbol{u}_{k-1}^{\top}
\end{bmatrix}^{\top} \}
^{\dagger}
\begin{bmatrix}
\boldsymbol{z}_{k-1}^{\top} & \boldsymbol{u}_{k-1}^{\top}
\end{bmatrix}^{\top}
].$

\emph{Proof:} Computing the gradient of (\ref{eq:obj}), we have,
\begin{eqnarray}
    \frac{\partial J_k}{\partial \boldsymbol{\Delta A}_k} &=&-\boldsymbol{y}_k^{\top} \boldsymbol{z}_{k-1}, \frac{\partial^2 J_k}{\partial \boldsymbol{\Delta A}_k^2} = \boldsymbol{z}_{k-1}^{\top} \boldsymbol{z}_{k-1},\nonumber\\
    \frac{\partial J_k}{\partial \boldsymbol{\Delta B}_k} &=&-\boldsymbol{y}_k^{\top} \boldsymbol{u}_{k-1}, \frac{\partial^2 J_k}{\partial \boldsymbol{\Delta B}_k^2} = \boldsymbol{u}_{k-1}^{\top} \boldsymbol{u}_{k-1}.
\end{eqnarray}
From the above gradient equations, the condition for minimizing the loss function $J_k$ is $\boldsymbol{y}_k=\boldsymbol{\Delta z}_k - \boldsymbol{\Delta A}_k \boldsymbol{z}_{k-1} - \boldsymbol{\Delta B}_k \boldsymbol{u}_{k-1}=0$, or equivalently $ \begin{bmatrix}
        \boldsymbol{\Delta A}_k & \boldsymbol{\Delta B}_k
    \end{bmatrix}
    = \boldsymbol{\Delta z}_k
    \left(
    \begin{bmatrix}
    \boldsymbol{z}_{k-1}^{\top} & \boldsymbol{u}_{k-1}^{\top}
    \end{bmatrix}^{\top}
    \right)
    ^{\dagger}$,
which is the proposed update law. Hence, the proposed update law minimizes the objective function (\ref{eq:obj}), which implies that at each timestep $k$, linear model (\ref{eq:uncertain_koop}) is updated such that the prediction error $\boldsymbol{\Delta z}_k$ is driven to the origin. $\qed$


In practice, it is more stable and robust to formulate the updates $\boldsymbol{\Delta A}_k$ and $\boldsymbol{\Delta B}_k$ based on error over a window of $w$ latest measurements, with more weightage ascribed to the most recent measurement rather than older data within this history stack. This can be incorporated by including exponentially decaying forgetting factor $\nu \in (0,1]$, which assigns diminishing importance to past data over the window. Hence, from (\ref{eq:pseudo}) we have, $\begin{bmatrix}
        \boldsymbol{\Delta A}_k & \boldsymbol{\Delta B}_k
    \end{bmatrix}
    = \boldsymbol{\Delta Z}_k
    \left(
    \begin{bmatrix}
    \boldsymbol{Z}_{k-1}^{\top} &  \boldsymbol{U}_{k-1}^{\top}
    \end{bmatrix}^{\top}
    \right)
    ^{\dagger},\,\text{where}$
$\boldsymbol{\Delta Z}_{k} = [\nu^{w-1} \boldsymbol{\Delta z}_{k-w+1}, \nu^{w-2}\boldsymbol{\Delta z}_{k-w},...,\boldsymbol{\Delta z}_{k}]$, $\boldsymbol{Z}_{k-1} = [\boldsymbol{z}_{k-w},\boldsymbol{z}_{k-w-1},...,\boldsymbol{z}_{k-1}]$ and $\boldsymbol{U}_{k} = [\boldsymbol{u}_{k-w},\boldsymbol{u}_{k-w-1},...,\boldsymbol{u}_{k-1}]$.

One key advantage of the proposed approach is the separation of the adaptation mechanism from the control algorithm, enabling the use of any controller suited to the application. In this study, Model Predictive Control (MPC) is chosen for its optimal control benefits and its ability to enforce state and input constraints. The controller is designed to integrate with the linear Koopman model. The following optimization problem is solved over the horizon of length $b$ using the Operator Splitting Quadratic Program (OSQP) solver:
\begin{eqnarray}
     \min_{Z,U} \sum_{k=0}^{b-1} & \Big( \left( \boldsymbol{C}z_k - x_{\text{ref,$k$}} \right)^{\top} P \left( \boldsymbol{C}z_k - x_{\text{ref,$k$}} \right) + \left (u_k \right)^{\top} R u_k \Big)\nonumber\\
     & + \left( \boldsymbol{C}z_b - x_{\text{ref,$b$}} \right)^{\top} P_b \left( \boldsymbol{C}z_b - x_{\text{ref,$b$}} \right)
    \label{MPC controller}
 \end{eqnarray}
 subject to: $\boldsymbol{z}_{k+1} = \hat{\boldsymbol{A}}_k\boldsymbol{z}_k + \hat{\boldsymbol{B}}_k\boldsymbol{u}_k, \, z_0 = \phi(x_0),\, \hat{\boldsymbol{A}}_0={\boldsymbol{A}},\,\hat{\boldsymbol{B}}_0={\boldsymbol{B}},\, x_{\text{min}} \leq \boldsymbol{C}\boldsymbol{z}_k \leq x_{\text{max}}, \, u_{\text{min}} \leq \boldsymbol{u}_k \leq u_{\text{max}}$, where the system parameters $\hat{\boldsymbol{A}}_k$ and $\hat{\boldsymbol{B}}_k$ are obtained via event-triggered updates discussed below, $x_{\text{ref},k}$ refers to the reference trajectory, $P$, $P_b$, and $R$ represent the weighing matrices. $[x_{min}, x_{max}]$ and $[u_{min}, u_{max}]$ represent constraint bounds on state and input respectively.

\section{Event Triggered Adaptation and Control}\label{sec3}
\begin{figure*}
   \centering  \includegraphics[width=0.85\textwidth]
  {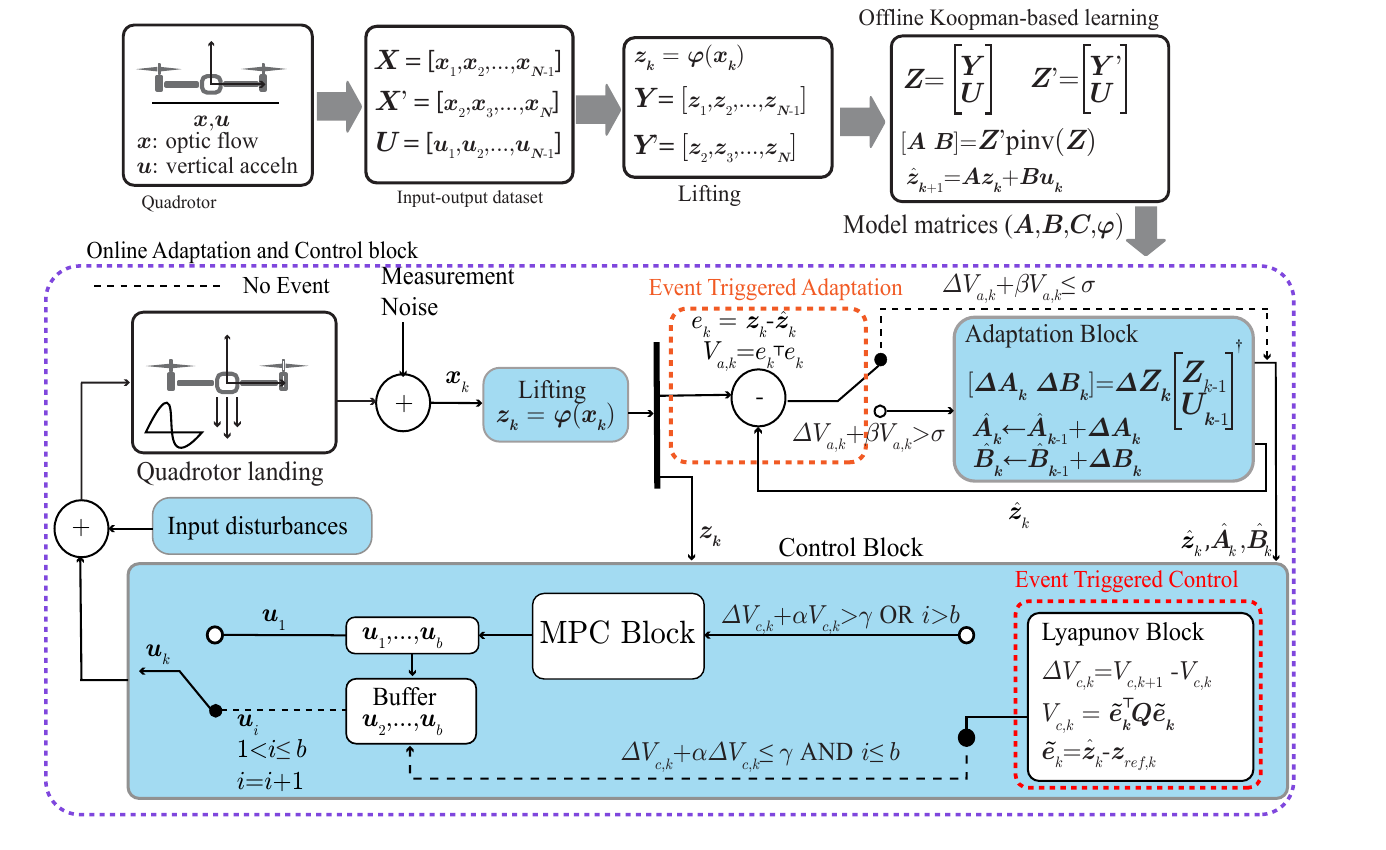}
  \caption{\textbf{Block diagram for Event-Triggered Model Adaptation and Event-Triggered Control}: In this framework, model and control updates occur aperiodically. The online adaptive architecture gets triggered when (\ref{Adaptation CLF}) is satisfied, and the MPC recomputes the control input only when $ev = 1$ is satisfied in (\ref{Event Trigger eq}).}  \label{Event_Triggered}
\end{figure*}
This section introduces the concept of Event-Triggered Adaptation and Event-Triggered Control. We design separate event triggers for model updates and control updates.



\subsection{Event Triggering Mechanisms}
 Asymptotic convergence of a discrete system is guaranteed if the Control Lyapunov Function (CLF) $V(z_k)$ satisfies the inequality $\inf_{u_k \in U} \left[ V(z_{k+1}) - V(z_k) \right] \leq -\Omega(V(z_k))$
where \( \Omega \) is a class \( \mathcal{K}_{\infty} \) function. To guarantee uniform ultimate boundedness of the prediction/tracking error, we introduce a threshold $\delta > 0$ such that:
\begin{equation}
    \inf_{u_k \in U} \left[ V(z_{k+1}) - V(z_k) \right] +\Omega(V(z_k)) \leq \delta,
    \label{Modified CLF}
\end{equation}


This formulation ensures that the closed-loop system converges the prediction and tracking error to the neighborhood of origin. The CLF bounds the states, with convergence governed by $\delta$, reducing event-triggered updates while balancing stability and performance. Novel triggering conditions for adaptation and control input updates are now proposed.

\subsubsection{Event-Triggering Mechanism for Adaptation}
The triggering condition for the adaptation scheme is:
\begin{equation}
     \Delta V_a(e_k) +\beta V_a(e_k) > \sigma,
    \label{Adaptation CLF}
\end{equation}
where $\sigma>0$ is the threshold parameter and $\beta \in (0,1)$ defines the rate at which $V_a(e_k)$ must decay at each time step. Here $V_a$ is chosen to be $e_k^{\top}e_k$, with $e_k = x_{k} - \hat{x}_{k}$, where $\hat{x}_{k}$ is the prediction of the Koopman model (\ref{eq:Koopman_lin}) and $x_k$ is the observation from the real system. Using (\ref{Adaptation CLF}) as the event trigger, the model parameters are updated at the triggering instant as $\hat{\boldsymbol{A}}_k = \hat{\boldsymbol{A}}_{k-1} + \boldsymbol{\Delta A}_k$ and $\hat{\boldsymbol{B}}_k = \hat{\boldsymbol{B}}_{k-1} + \boldsymbol{\Delta B}_k$ using the update law (\ref{eq:pseudo}). Note that the parameters $\boldsymbol{A}$, $\boldsymbol{B}$, $\boldsymbol{\Delta A}_k$, and $\boldsymbol{\Delta B}_k$ (thus $\hat{\boldsymbol{A}}_k$, $\hat{\boldsymbol{B}}_k$) are always bounded because of the EDMD algorithm and the assumption that the dataset is i.i.d.

\subsubsection{Event-Triggering Mechanism for Control}
The Lyapunov function for system (\ref{eq:uncertain_koop}) is chosen to be :
\begin{equation}
     V_c(\tilde{\boldsymbol{e}}_k){=}  \tilde{\boldsymbol{e}}_k^{\top}Q \tilde{\boldsymbol{e}}_k,\,Q=Q^{\top}>0,\,\tilde{\boldsymbol{e}}_k{=}\boldsymbol{z}_k{-}\boldsymbol{z}_{ref,k}.
    \label{Control Lyapunov}
\end{equation}



The bounded CLF where $\alpha \in (0,1)$ and $\gamma>0$ is written as:
\begin{eqnarray}
\label{bounded_clf_control}
\Delta V_{c}(\tilde{\boldsymbol{e}}_k) +\alpha V_c(\tilde{\boldsymbol{e}}_k) \leq \gamma,
\end{eqnarray}
so that by invoking 
\begin{eqnarray}
\label{M_defn}
q(k) =
\begin{bmatrix}
\tilde{\boldsymbol{e}}_k \\
u_k
\end{bmatrix},
\quad
M_k =
\begin{bmatrix}
\hat{A}_k^{\top}Q\hat{A}_k-Q & \hat{A}_k^{\top}Q\hat{B}_k \\
\hat{B}_k^{\top}Q\hat{A}_k & \hat{B}_k^{\top}Q\hat{B}_k,
\end{bmatrix},
\end{eqnarray}
 we have from (\ref{bounded_clf_control}),
\begin{align}
    \varepsilon_k {:=} & \; q(k)^{\top} M_k q(k) {+} \alpha V_c (\tilde{\boldsymbol{e}}_k) {+} 2\tilde{\boldsymbol{e}}_k^{\top} \hat{A}_k^{\top} Q(\hat{A}_k z_{ref,k} - z_{ref,k}) \nonumber \\
    & {+} 2\boldsymbol{u}_k^{\top} \hat{B}_k^{\top} Q(\hat{A}_k z_{ref,k} - z_{ref,k}) \nonumber \\
    & {+} (\hat{A}_k z_{ref,k} - z_{ref,k})^{\top} Q(\hat{A}_k z_{ref,k} - z_{ref,k}) \leq \gamma.
    \label{Control Trigger}
\end{align}

Finally, the event mechanism for control update is shown as follows:
\begin{equation}
ev = 
\begin{cases} 
1, & \text{if } \varepsilon_k {>} \gamma \text{ or } i {>} b \\
0, & \text{otherwise},
\end{cases}
\label{Event Trigger eq}
\end{equation}
where $i$ represents the number of consecutive instances where the MPC has not been triggered. When an event is triggered $(ev = 1)$, the input sequence over the horizon is computed using (\ref{MPC controller}), and only the first input $u_1$ is applied to the system, as shown in Fig.\ref{Event_Triggered}. If no event is triggered $(ev = 0)$, the control input from the last stored sequence is applied sequentially until either (\ref{Control Trigger}) is violated or the sequence is exhausted ($i=b$). Thus, we have:
\begin{equation}
u_k = 
\begin{cases} 
\text{Solution of (\ref{MPC controller}}), & \text{if } ev = 1 \\
U^{*}(i|k_{ev}), & \text{otherwise}
\end{cases}
\label{input}
\end{equation}
where $k_{ev}$ is the time instant at which the most recent event has occurred and $U^{*}(i|k_{ev})$ the control sequence calculated at that instant using (\ref{MPC controller}). 


Fig.~\ref{Event_Triggered} illustrates the proposed event-triggered adaptation and control scheme. A monocular camera captures the optic flow, which is compared to the flow predicted by the linear Koopman model.  If (\ref{Adaptation CLF}) is satisfied at the time step $k$, the system matrices are updated according to (\ref{eq:pseudo}). This reduces computations by addressing uncertainties only when needed. The updated model is used in the MPC block to calculate control inputs, with optimization triggered either by exceeding a control Lyapunov threshold $\gamma$ or after $b$ steps without a new event. This strategy preserves performance while again reducing the computational overhead.

\subsection{Convergence Analysis}
\emph{Theorem 1: If the linear system (\ref{eq:Koopman_lin}) is updated using (\ref{eq:pseudo}) and the optimization objective for MPC is defined as in (\ref{MPC controller}) with the triggering mechanism given by (\ref{Adaptation CLF}) and (\ref{Event Trigger eq}), then the prediction error $e_k$ and tracking error $\bar{e}_k$ achieve global convergence to the uniform ultimate bound $(\sqrt{{\sigma}/{\beta}})$ and $(\sqrt{\frac{\sigma}{\beta}} + \sqrt{\frac{\gamma}{\alpha \lambda_{min}(Q)}})$ respectively under the event-triggering rule (\ref{Adaptation CLF}) and (\ref{Control Trigger}).Furthermore, there exists a positive constant $\tau^{*}$ such that $\tau_k \geq \tau^{*} \forall k \in \mathbb{N}$ where the inter-execution time can be defined as $\tau_k = t_{ev+1} - t_{ev}$.}

\emph{Proof:}

Convergence of the Prediction Error:
Between triggering instants, it follows from (\ref{Adaptation CLF}),
\begin{eqnarray}
V_{a,k+1} &\leq& (1{-}\beta)V_{a,k} {+} \sigma, \nonumber \text{ likewise,}\\ 
V_{a,k+M} &\leq& (1{-}\beta)^M V_{a,k} {+} (\sigma/\beta) ( 1 - (1{-}\beta)^M ),
\end{eqnarray}
so that, with $0<\beta<1$, as \( M\) becomes large, we have
\[
V_{a,k+M} \leq \sigma/\beta, \quad |e_{k+M}| \leq \sqrt{\sigma/\beta}.
\]
Hence, the prediction error has the upper bound $\sqrt{{\sigma}/{\beta}}$.

Convergence of the Tracking Error:
Now consider the tracking error \( \bar{e}_k = z_k - z_{\text{ref},k} \), which can be decomposed as:
\(
\bar{e}_k = (z_k - \hat{z}_k) + (\hat{z}_k - z_{\text{ref},k}) = e_k + \tilde{e}_k,
\)
where \( \tilde{e}_k = \hat{z}_k - z_{\text{ref},k} \).  
The convergence proof for time-triggered MPC follows directly from \cite{sun2023event} and is thus excluded in the interest of brevity. However, in between events, when \( ev = 0 \) in (\ref{Event Trigger eq}), we have \( V_{c,k+i} \leq \gamma/\alpha \) for large \( i \). Therefore, \( V_c \) is bounded above by \( \gamma/\alpha \), implying that the $\|\tilde{e}_k\|$ is bounded above by \( \sqrt{\gamma/\alpha \lambda_{min}\{Q\}} \). Combining both errors,
\(
\bar{e}_{k+M} = (z_{k+M} - \hat{z}_{k+M}) + (\hat{z}_{k+M} - z_{\text{ref},k}),
\)
\[
\|\bar{e}_{k+M}\| \leq \|z_{k+M} - \hat{z}_{k+M}\| + \|\hat{z}_{k+M} - z_{\text{ref},k}\|,
\]
\[
\|\bar{e}_{k+M}\| \leq \sqrt{\frac{\sigma}{\beta}} + \sqrt{\frac{\gamma}{\alpha \lambda_{\min}(Q)}}.
\]
Therefore, the tracking error \( \bar{e}_k \) is uniformly bounded.

We now demonstrate that the event-triggered control scheme guarantees Zeno-free behavior. Specifically, we show that there exists a minimum inter-event time \( \tau^* \) such that \( \tau_k \geq \tau^* \) for all \( k \in \mathbb{N} \). We need to show that $V_k = \Delta V_c(\tilde{\boldsymbol{e}}_k) + \alpha V_c(\tilde{\boldsymbol{e}}_k) $ does not grow too quickly after each triggering event, that is, there exists a minimum number of steps before $V_k$ exceeds \(\gamma\) again.
Now, noting that \( \tilde{\boldsymbol{e}}_{k+1} - \tilde{\boldsymbol{e}}_k = \hat{\boldsymbol{z}}_{k+1} - \hat{\boldsymbol{z}}_k \), we have
\(
\|\hat{\boldsymbol{z}}_{k+1} - \hat{\boldsymbol{z}}_k\| = \|\hat{\boldsymbol{A}}_k \hat{\boldsymbol{z}}_k + \hat{\boldsymbol{B}}_k u_k - \hat{\boldsymbol{z}}_k\|.
\)
Let \( M_A = \|\hat{\boldsymbol{A}}_k - I\| \) and \( M_B = \|\hat{\boldsymbol{B}}_k\| \), where \( I \) is the identity matrix. Thus, we obtain:\(
    \|\hat{\boldsymbol{z}}_{k+1} - \hat{\boldsymbol{z}}_k\| \leq M_A \|\hat{\boldsymbol{z}}_k\| + M_B |u_k|.
\)
From (\ref{MPC controller}), we can define \(
L_1 = max\left\{ |x_{min}|, |x_{max}| \right\}.
\) and \( L_2 = max\left\{ |u_{min}|, |u_{max}| \right\}.
\) Since \( \hat{z}_k \) is composed of \( x \) and \( x^2 \), the norm \( \hat{z}_k  \) can be bounded as:\(
\| \hat{z}_k \| = \sqrt{x^2 + (x^2)^2} \leq \sqrt{L_1^2 + L_1^4}.\)
Thus, the upper bound on \( \| \hat{z}_k \| \), denoted as \( z_{max} \), can be explicitly defined using \( L_1 \) as
\(z_{max} = \sqrt{L_1^2 + L_1^4}.\)
Since \( V_c \) is quadratic and \(
\| \tilde{\boldsymbol{e}}_k \| = \| \hat{\boldsymbol{z}}_k - \boldsymbol{z}_{\text{ref}} \| \leq \| \hat{\boldsymbol{z}}_k \| + \| \boldsymbol{z}_{ref} \| \leq 2  \boldsymbol{z}_{max},
\) 
it follows that \( V_c \) is Lipschitz continuous with Lipschitz constant
\(
L_v = 4 \lambda_{max}\{Q\} \boldsymbol{z}_{max}.
\)
Thus, \[
|V_{k+1} - V_k| \leq L_v \|z_{k+2} - z_{k+1}\| + L_v \|z_{k+1} - z_{k}\| + \alpha L_v \|z_{k+1} - z_k\|,
\]
\begin{equation}
|V_{k+1} - V_k| \leq L_v \overline{M}(2+\alpha).
\label{Zeno}
\end{equation}
where, $\overline{M} = M_A\boldsymbol{z}_{max} + M_B L_2$. The event is triggered when $V_k > \gamma$, and the states evolve for $V$ to reach this threshold again. This allows us to establish a lower bound on the minimum number of steps $j$ required to trigger the next event, which can be directly obtained from (\ref{Zeno}) as $j \geq {\gamma}/({L_v \overline{M}(2+\alpha)}).$ This implies that after an event is triggered, it will take at least \( j \) steps before the next event can occur, thereby guaranteeing Zeno-free behavior.

The above proof demonstrates that the prediction and tracking errors are globally bounded, and the event-triggered mechanism is Zeno-free, ensuring a positive minimum inter-event time. $\qed$

\begin{figure*}[ht!]
     \centering
     \includegraphics[width=0.9\textwidth]{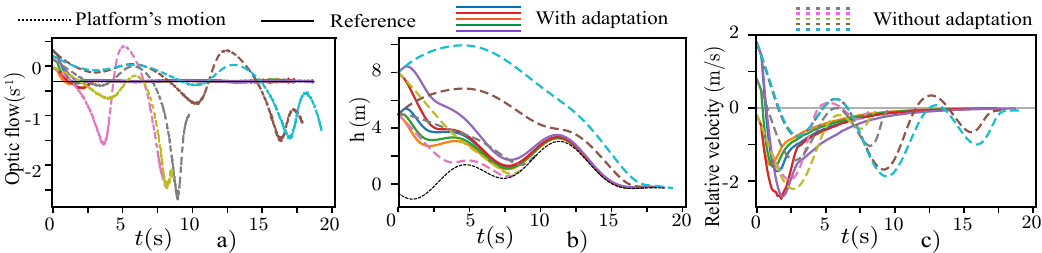}
     \caption{Simulation results for landing on a vertically oscillating platform for different initial conditions ($h(0) \in \{5, 8\} \, m$, $v(0) \in \{1, 0, -1\} \,m/s$) for both ETAC and non-adaptive ETC algorithms. a) Optic flow tracking $(x_{ref} = -0.3s^{-1})$ b) Absolute height of UAV and the platform c) Relative velocity between the UAV and the platform.
}
    \label{fig:event_all}
\end{figure*}

\section{Simulation Results}\label{sec5}
This section presents simulation results demonstrating the adaptive Koopman framework's performance in vertical UAV landings on an oscillating platform. The offline linear model (\ref{eq:Koopman_lin}) is learned using $100$ trajectories with $150$ data points each generated from simulation of the vertical motion of the UAV, and the observable functions $\varphi(x) = \{x, x^2\}$ are used for lifting. To assess the robustness and generalizability of the proposed algorithm, a series of simulations are conducted from an altitude of $\{5, 8\} \, m$ with an initial velocity of $\{1, 0, -1\} \,m/s$. The thresholds $\sigma$ and $\gamma$ are chosen as $5\times10^{-8}$ and $10^{-4}$ respectively, with $\alpha=\beta=0.09$ and $Q=100 I$. In addition, we evaluate the effectiveness of our approach by comparing the Event-Triggered Adaptation and Control (ETAC) with Event-Triggered Control (ETC) with no model update capability (no adaptation) and Time-Triggered Adaptation and Control (TTAC). The simulations were conducted with a $12^{th}$ Gen Intel(R) Core(TM) i7-12700 processor. The ground effect is incorporated into the simulations as described in \cite{sanchez2017characterization}. To enhance generalizability, the platform's motion is modeled stochastically as the sum of $10$ sinusoidal signals, where each signal is described by $ h_p(t) = h_{p0} + h_{p0} \sin(\omega_m t + \theta_m)$. Here, \( h_p(t) \) represents the height for each signal, with $h_{p0}=0.5 m$ as the amplitude,  the angular frequency randomly selected from the range $\omega_m \in [0.1,1] rad/s$, and $\theta_m \in [0,2\pi] rad$ for each signal.



To evaluate the impact of sensor noise, simulations are conducted with optic flow measurements corrupted by a signal-to-noise ratio (SNR) of 35 dB. The simulation results for landing on a moving platform with and without adaptation using event-triggered updates are illustrated in Fig. \ref{fig:event_all}. Without adaptation, the optic flow trajectory (as shown by the dotted line) diverges as the UAV approaches the ground. In contrast, the adaptive approach enables optic flow convergence to the desired reference value of $-0.3s^{-1}$. The inability of the non-adaptive algorithm to address uncertainties arising due to ground effect and the platform motion leads to erroneous control signals, preventing a smooth touchdown, as seen from Figs. \ref{fig:event_all}b, \ref{fig:event_all}c. In contrast, the adaptive framework effectively accounts for these uncertainties by continuously updating the Koopman model online, thus ensuring accurate tracking, which results in a precise soft landing maneuver for all the initial conditions. These results demonstrate the efficacy of the proposed algorithm in achieving a smooth landing on an oscillating platform by relying solely on noisy optic flow. 

To quantitatively assess the performance of time- and event-triggered methods, a performance comparison study is undertaken in terms of RMS error computed over the last $4s$ of the trajectory, total control effort $(= \int_{0}^{T} |u(t)|dt)$, average computation time per iteration, and the total number of adaptation and control events, with the results summarized in Table \ref{tab:noise_comp}. Event-triggered updates achieve comparable tracking performance while avoiding $325$ out of $1721$ potential control events and $821$ adaptation events, thus reducing the computational overhead by $33.3\%$. This leads to significantly reduced computation time, with only a marginal increase in RMSE relative to the TTAC scheme. These findings demonstrate that ETAC provides superior overall efficiency and performance compared to TTAC and ETC.

\begin{table}[h!]
\caption{\label{tab:noise_comp}Comparison of TTAC and ETAC for UAV landing on a vertically oscillating platform (initial height: $5m$, initial velocity: $1m/s$, $35dB$ measurement noise)}
\centering
\begin{tabular}{|p{5cm}|p{0.8cm} p{0.8cm}|}
\hline
{\textbf{Metric}} & \textbf{TTAC} & \textbf{ETAC} \\
\hline
No. of iterations  &  $1769$ & $1721$  \\
Avg. computation time per iteration (in $s$) & $0.00228$ & $0.00168$ \\
Total control effort (in ms$^{-2}$)  & $5.99$ & $5.82$ \\
RMSE of optic flow (in s$^{-1}$) & $0.0099$ & $0.011$ \\
Terminal time (in s)  & $17.69$ & $17.21$  \\
Terminal altitude (in m) & $ 0.05$ & $0.05$  \\
Terminal velocity (in m/s)  & $-0.015$ & $-0.01$   \\
Adaptation events  & $1769$ & $900$ \\
Control events  & $1769$ & $1396$ \\
Total events avoided & $-$ & $1146$ \\
\hline
\end{tabular}
\end{table}

\section{Conclusion}\label{sec6}
This paper addresses the problem of achieving a smooth landing on a moving platform with only a monocular camera to measure the optic flow output. The paper presents a Koopman-based data-driven strategy that learns a linear model of the optic flow dynamics induced by vertical motion. The proposed strategy incorporates a novel adaptation framework that updates the parameters of the linear Koopman model online, which enables it to adapt to uncertainties due to the presence of ground effect and landing platform motion. In addition, to further reduce the computational overhead, a novel event-triggered adaptation and control scheme is proposed with independent mechanisms to trigger the adaptation and control loops. The efficacy and robustness of the proposed algorithms are showcased through simulation results demonstrating a smooth landing on a moving platform in the presence of sensor noise. Future work involves experimentally implementing the proposed Event-Based Adaptive Koopman framework in a hardware-in-the-loop setup.




\bibliographystyle{IEEEtran.bst}
\bibliography{References}

\end{document}